\documentclass[10pt]{article}
\usepackage{amsmath}
\usepackage{amsfonts}
\usepackage{amssymb}
\usepackage{graphicx}
\def\be{\begin{equation}}
\def\ee{\end{equation}}
\def\bea{\begin{eqnarray}}
\def\eea{\end{eqnarray}}
\title{On the smoothness of static multi-black hole solutions of higher-dimensional Einstein-Maxwell theory}

\author{Graeme N. Candlish and Harvey S. Reall \\ School of Physics and Astronomy, University of Nottingham \\ Nottingham NG7 2RD, UK}

\begin{document}

\maketitle

\begin{abstract}
Previous work has shown that static multi-black hole solutions of higher-dimensional Einstein-Maxwell theory do not possess smooth horizons. We show that the lack of smoothness is worse than previously demonstrated. We consider solutions describing multiple black holes on a common axis. In five dimensions, the metric is generically twice, but not three times, continuously differentiable at the horizon. 
The Maxwell field is generically continuous, but not differentiable, at the horizon. In more than five dimensions, the metric is once, but not twice, continuously differentiable, and there is a parallely-propagated curvature singularity at the horizon. The Maxwell field strength is again continuous, but not differentiable, at the horizon. 
\end{abstract}

\section{Introduction}
It has long been known that four-dimensional Einstein-Maxwell theory admits static multi-black hole solutions. The existence of these solutions can be attributed to a balance of electromagnetic and gravitational forces. The (degenerate) horizon of each black hole in these solutions is known to be analytic \cite{Hawking}.

String theory motivates the study of multi-black hole solutions of higher-dimensional Einstein-Maxwell theory. Perhaps surprisingly, it turns out the horizons of such solutions are generically not smooth. This was first suggested by Gibbons {\it et al} \cite{Gibbons} and later proved by Welch \cite{Welch95} for the case in which the black holes lie on a common axis. Welch considered a timelike geodesic along the axis as it approaches the horizon of one of the black holes. He calculated the components of the Riemann tensor in a parallely propagated frame. For a two-black hole solution, he showed that second derivatives (with respect to proper time) of some of these components diverge at the horizon. This proves that the Riemann tensor is not $C^2$ (twice continuously differentiable) at the horizon and hence the metric is not $C^4$ at the horizon. Below we shall present a much shorter proof of this result.

Stating things more precisely, the work of Welch proves that there does not exist a $C^4$ extension of the metric through the horizon. However, the question still remains: how smooth is the solution? To answer this question, it is necessary to construct a coordinate chart that covers the horizon. One such chart was constructed in \cite{Myers}. The resulting metric is $C^0$ but not $C^1$. Hence previous work has demonstrated that there exists a $C^0$ extension, but not a $C^4$ extension.
The aim of the present paper is to sharpen these results.

To our surprise, it turns out that higher-dimensional multi-black hole solutions are even less smooth than proved by Welch. By considering {\it non-axial} geodesics, we shall show that the metric at the horizon of a two-black hole solution is not $C^3$ in five dimensions.
We shall then introduce Gaussian null coordinates to construct a coordinate chart in which the metric is $C^2$ in a neighbourhood of the horizon. Hence the metric is $C^2$, but not $C^3$. The Maxwell field strength is $C^0$, but not $C^1$, at the horizon. 

In higher dimensions, things are even worse: the metric is $C^1$, but not $C^2$, in more than five dimensions. In fact, we shall show that there is a parallely propagated curvature singularity at the horizon. This sounds problematic: one usually demands that the metric should be $C^2$ in order that the Einstein equation makes sense. However, the Maxwell tensor is still $C^0$ and hence the RHS of the Einstein equation is $C^0$, so the Einstein tensor must be $C^0$ even though there is a curvature singularity at the horizon. Therefore the Einstein equation is still satisfied at the horizon.

What lies behind the horizon? When dealing with a non-smooth horizon, this question need not admit a unique answer. We shall show that there are infinitely many different ways of matching the solution across the horizon onto an interior solution, whilst preserving the degree of differentiability just discussed. This phenomenon also occurs for four-dimensional cosmological multi-black hole solutions \cite{Brill94}. 

As in the cosmological case \cite{Brill94}, the degree of differentiability at the horizon of a given black hole can be increased by introducing extra black holes in an appropriate (finely-tuned) manner \cite{Welch95}. One can even make the horizon of one black hole analytic. However, the horizons of the other black holes will still be non-smooth unless one considers an infinite array of black holes. Surprisingly, this "smoothness enhancement" is only possible in $d=5$: there are no $d>5$ solutions with a metric smoother than $C^1$ at the horizon.

This paper is organized as follows. In section \ref{start}, we describe the general multi-black hole solution with which we will be working and present a brief argument that demonstrates that the horizon of a two-black hole solution is not $C^4$. In section \ref{5dmulticentre} we analyze $d=5$ multi-black hole solutions in detail. The $d>5$ solutions are discussed in section \ref{higherdsmoothness}. Some technical details are given in the Appendices.

\section{Multi-centre co-axial solutions}
\label{start}
The solutions we are considering in this paper are those of $d$-dimensional Einstein-Maxwell theory, with action:
\begin{equation}
\int d^d x \sqrt{-g} \left( R - \left( \frac{d-2}{8(d-3)} \right) F_{\mu \nu} F^{\mu \nu} \right)
\end{equation}
The static multi-centre solutions to this theory have the following metric:
\begin{equation}
\label{metric}
ds^2 = -H^{-2} dt^2 + H^{2/(d-3)}(dr^2 + r^2 d\theta^2 + r^2 \sin^2 \theta d\Omega_{d-3}^2)
\end{equation}
where $d\Omega_{d-3}^2$ denotes the metric on the unit $(d-3)$-sphere. The Maxwell field strength is $F=dA$, where
\begin{equation}
\label{maxwell}
 A = - H^{-1} dt.
\end{equation}
The function $H$ is harmonic on $R^{d-1}$, with poles corresponding to the locations of the event horizons of the black holes. We have chosen our coordinate system in the metric above such that the harmonic function for the multi-centre co-axial solution depends only on $r$ and $\theta$. Thus the multi-centre harmonic function for a line of $N$ black holes in our case is:
\begin{equation}
\label{harmonic}
H = 1 + \frac{\mu_1}{r^{d-3}} + \sum_{i=2}^N \frac{\mu_i}{(r^2 + a_i^2 - 2a_i r\cos \theta)^{(d-3)/2}}
\end{equation}
where the $a_i$ denotes the position along this axis of the $i$-th black hole, and $\mu_i$ is a mass parameter. We shall investigate smoothness of the horizon at $r=0$. To this end, it is convenient to expand $H$ in (axisymmetric) spherical harmonics:
\begin{equation}
H = \frac{\mu_1}{r^{d-3}} + \sum_{n=0}^{\infty} h_n r^n Y_n(\cos \theta),
\end{equation}
where the harmonics $Y_n(\cos \theta)$ are given by Gegenbauer polynomials:
\be
Y_n(\cos \theta) = C_n^{\frac{d-3}{2}} (\cos \theta)
\ee
with the first few being
\begin{equation}
\begin{split}
C_0^{\frac{d-3}{2}} &= 1 \\
C_1^{\frac{d-3}{2}} &= (d-3) \cos \theta \\
C_2^{\frac{d-3}{2}} &= \frac{(d-3)}{2}((d-1) \cos^2 \theta - 1).
\end{split}
\end{equation}
The coefficients $h_n$ are given by
\be
\label{hnalld}
h_n = \delta_{n,0} + \sum_{i=2}^{N} \frac{\mu_i}{|a_i|^{d-3} a_i^n}.
\ee
For generality, we shall take $h_n$ to be independent parameters until we need to specify to particular examples.

\subsection{Short argument for non-smoothness}

\label{susy}

The argument of \cite{Welch95} involved calculating derivatives of components of the Riemann tensor in a parallely propagated frame. Here we shall present a much shorter argument that allows one to reach the same conclusion, namely that the metric is not $C^4$ at the horizon of a two-black hole solution.

We start from the assumption that the metric admits a $C^k$ extension through the horizon of one of the holes, with $k \ge 2$, and that this extension admits a $C^1$ Killing vector field $V$ which coincides with $\partial/\partial t$ outside the horizon.\footnote{
An alternative set of assumptions for the $d=5$ case is motivated by supersymmetry. The solution (\ref{metric},\ref{maxwell}) admits a super-covariantly constant spinor $\epsilon$, from which one  can construct a Killing vector field $V_a \sim \bar{\epsilon} \gamma_a \epsilon$ \cite{gkltt} which coincides with $\partial/\partial t$. We could assume that the extension through the horizon preserves supersymmetry, i.e., $\epsilon$ extends through the horizon. If the Maxwell field is $C^0$ (and the metric assumed $C^k$, $k \ge 2$ as in the main text) then $\epsilon$ will be $C^1$ and hence so will $V$.}
Note that $C^1$ is the minimal assumption for Killing's equation to make sense. However, we can now show\footnote{We thank P. Chrusciel for this argument.} that $V$ must actually be $C^k$. As $V$ is a Killing vector, we have the following relation:
\begin{equation}
\label{killing}
\nabla_a \nabla_b V_c = -R_{bca}^d V_d
\end{equation}
where $R_{bca}^d$ is the Riemann tensor. As the metric is $C^k$, the Riemann tensor must be $C^{k-2}$, so at least $C^0$ as $k \ge 2$. Hence the RHS of (\ref{killing}) is $C^0$ and therefore so must be the LHS, which implies that $V$ must be $C^2$. Plugging this back into (\ref{killing}), the right hand side must now be $C^{\text{min}(2,k-2)}$, so by the previous argument the Killing vector must be $C^{\text{min}(2,k-2)+2} = C^{\text{min}(4,k)}$. Repeating this argument we will eventually learn that $V$ is $C^{\text{min}(n,k)}$ where $n > k$, so by induction we arrive at the conclusion that $V$ must be $C^k$.

Now consider the norm of $V$:
\begin{equation}
\label{relation}
  V_{\mu} V^{\mu}=-H^{-2}
\end{equation}
Since both the metric and $V$ are $C^k$, it follows that $H^{-2}$ must be $C^k$ at the horizon. We can use this to determine an upper bound on $k$ by considering the differentiability of $H^{-2}$ along axial null geodesics.

Instead of the spherical polar coordinates of (\ref{metric}), we can use cylindrical polar coordinates $(\rho,z,\Omega_{d-3})$ so that the metric has the form
\be
 ds^2 = -H^{-2} dt^2 + H^{2/(d-3)} \left( d\rho^2 + dz^2 + \rho^2 d\Omega_{d-3}^2 \right).
\ee
For a two-centre solution we have
\be
 H = 1 + \frac{\mu_1}{(\rho^2 + z^2)^{(d-3)/2}} + \frac{\mu_2} {(\rho^2 + (z-a)^2)^{(d-3)/2}}.
\ee
Consider a future-directed null geodesic approaching the origin along the positive $z$-axis. Such a geodesic has $\rho=0$. From energy conservation we have (for appropriately normalized affine parameter $\lambda$)
\be
 \frac{dt}{d\lambda} = H^2
\ee
and from the null condition we have
\be
 \frac{dz}{d\lambda} = - H^{\frac{d-4}{d-3}} = - \mu_1^{\frac{d-4}{d-3}} z^{4-d} \left( 1 + \frac{z^{d-3}}{\mu_1} + \frac{z^{d-3} \mu_2}{\mu_1 (a-z)^{d-3}} \right)^{\frac{d-4}{d-3}}.
\ee 
Solving this equation for small $z$ (with $z \rightarrow 0+$ as $\lambda \rightarrow 0-$) gives
\begin{multline}
z(\lambda) = (d-3)^{1/(d-3)} \mu_1^{\frac{d-4}{(d-3)^2}}(-\lambda)^{1/(d-3)} \\ + (\frac{1}{2}d-2)(d-3)^{\frac{4-d}{d-3}}\mu_1^{-1/(d-3)^2}(1+a^{3-d}\mu_2)(-\lambda)^{\frac{d-2}{d-3}} \\ + \frac{d-4}{2d-5}(d-3)^{\frac{d-1}{d-3}}\mu_1^{\frac{d-5}{(d-3)^2}}\mu_2a^{2-d}(-\lambda)^{\frac{d-1}{d-3}} + \mathcal{O}((-\lambda)^{\frac{d}{d-3}})
\end{multline}
This implies that, along this geodesic we have
\begin{multline}
H^{-2} = (d-3)^2\mu_1^{-2/(d-3)}\lambda^2 + (d-3)^2(d-2)\mu_1^{-2/(d-3)}(1+a^{3-d}\mu_2)\lambda^3 \\ - \frac{2(d-1)(d-3)^{\frac{4d-11}{d-3}}}{2d-5}\mu_1^{\frac{5-2d}{(d-3)^2}}\mu_2a^{2-d}(-\lambda)^{\frac{3d-8}{d-3}} + \mathcal{O}((-\lambda)^{\frac{3d-7}{d-3}})
\end{multline}
This is an expansion in powers of $\lambda^{1/(d-3)}$. For $d=4$, such an expansion is analytic in $\lambda$, as expected from \cite{Hawking}. However, if $d>4$ then the third term is not $C^4$ at $\lambda=0$. Hence $H^{-2}$ is not $C^4$ at the horizon so we must have $k < 4$, i.e., the metric is not $C^4$, which was the conclusion of Welch \cite{Welch95}. In later sections, we shall consider the behaviour of invariants along {\it non-axial} geodesics and show that this leads to the conclusion that the metric is even less smooth.

\subsection{Gaussian null coordinates}
\label{gauss}

In later sections we will employ Gaussian null coordinates to extend the metric through the horizon. In this section we give a brief description of how such coordinates are constructed. More details can be found in \cite{Wald}. We shall restrict attention to extensions that preserve the $R \times SO(d-3)$ symmetry of the metric (\ref{metric}). An argument of the previous section shows that the associated Killing fields will have the same degree of smoothness as the metric.

Consider a single component of the event horizon ${\cal H}^+_0$. The intersection $H_0$ of this with a spatial hypersurface is topologically a sphere $S^{d-2}$. Introduce coordinates $x^i$ on $S^{d-2}$. Let $V$ denote the generator of time translations ($V=\partial/\partial t$ in the coordinate of (\ref{metric})). $V$ is tangent to the null geodesic generators of ${\cal H}^+_0$ (this follows from the fact that $V$ generates a symmetry and is null on the horizon). Define a coordinate $v$ on ${\cal H}^+_0$ to be the parameter-distance from $H_0$ along integral curves of $V$. This defines a coordinate chart $(v,x^i)$ on a neighbourhood of $H_0$ in ${\cal H}^+_0$. 

Now let $U$ be the unique (past-directed) null vector field satisfying $U \cdot V = 1$ and $U \cdot \partial/\partial x^i=0$ on ${\cal H}^+_0$. Let $\gamma(v,x^i)$ be the null geodesic that starts at the point with coordinates $(v,x^i)$ in ${\cal H}^+_0$ and has tangent $U$ there. Gaussian null coordinates are defined in a neighbourhood of $H_0$ by ascribing coordinates $(v,\lambda,x^i)$ to the point affine parameter distance $\lambda$ from ${\cal H}_0^+$ along $\gamma(v,x^i)$.
The metric in these coordinates takes the form
\be
 ds^2 = -H(\lambda,x)^{-2} dv^2 + 2 dv d\lambda + 2 \lambda h_i (\lambda,x)dv dx^i + h_{ij}(\lambda,x) dx^i dx^j.
\ee
Note that $V=\partial/\partial v$ in these coordinates, and hence $g_{vv}$ is determined by $V^2 = -H^{-2}$. 

We can be more explicit about the coordinates $x^i$ on $S^{d-2}$. It is convenient to take $x^i=(\Theta,\hat{\Omega}_{d-3})$, where $(\Theta,\hat{\Omega}_{d-3})$ are the limiting values of $(\theta,\Omega_{d-3})$ along $\gamma$ as $\gamma$ approaches ${\cal H}^+_0$. The $SO(d-2)$ symmetry implies that angular momentum is conserved along $\gamma$. Hence, since we have initial condition $d\Omega_{d-3}/d\lambda = 0$ at $\lambda=0$, we must have $d\Omega_{d-3}/d\lambda \equiv 0$, i.e., $\Omega_{d-3}$ is constant along $\gamma$. Hence $\Omega_{d-3} \equiv \hat{\Omega}_{d-3}$. However, $\Theta$ and $\theta$ will not agree. In these coordinates, we will have
\be
\label{generalmetric}
ds^2 = -H^{-2} dv^2 + 2dvd\lambda + \lambda f_1(\lambda,\Theta)dvd\Theta +  f_2(\lambda,\Theta)d\Theta^2 + H^{2/(d-3)} r^2\sin^2 \theta d\Omega_{d-3}
\ee
where now $r=r(\lambda,\Theta)$ and $\theta=\theta(\lambda,\Theta)$ are determined once we know the geodesic $\gamma$.

The problem of bringing the metric to the Gaussian null form (\ref{generalmetric}) amounts to solving the geodesic equation to determine $\gamma(v,\Theta,\Omega_{d-3})$ using the original coordinate system of (\ref{metric}). As we have already seen, $\Omega_{d-3}$ is constant along $\gamma$. Energy conservation, together with $U \cdot V=1$ gives
\be
\label{energycons}
 \frac{dt}{d\lambda} = -H^2(r,\theta).
\ee
The null condition then reduces to
\be
\label{constr}
 0 = - H^2 + H^{2/(d-3)} \left( \dot{r}^2 + r^2 \dot{\theta}^2 \right),
\ee
where a dot denotes a derivative with respect to $\lambda$. The final equation we need is the geodesic equation for $r$:
\be
\label{req}
\ddot{r} - H^{\frac{d-5}{d-3}} \partial_r H + \frac{1}{d-3} H^{-1} \dot{r}^2 \partial_r H  - \frac{1}{d-3} H^{-1} r^2 \dot{\theta}^2 \partial_r H - r \dot{\theta}^2 + \frac{2}{d-3} H^{-1} \dot{r} \dot{\theta} \partial_{\theta} H =0.
\ee
The problem reduces to solving (\ref{constr}) and (\ref{req}) simultaneously to determine $r(\lambda,\Theta)$ and $\theta(\lambda,\Theta)$, where the initial conditions are
\be
\label{init}
 r(0,\Theta)=0, \qquad \theta(0,\Theta)=\Theta, \qquad \left(\frac{d \theta}{d\lambda}\right)_{\lambda=0} = 0.
\ee
Once $r$ and $\theta$ are known as functions of $\lambda$ and $\Theta$, equation (\ref{energycons}) can be integrated:
\be
\label{tsol}
 t = v -  T(\lambda,\Theta), \qquad T(\lambda,\Theta) \equiv \int H(\lambda,\Theta)^2 d\lambda,
\ee
with $v$ arising as the constant of integration.

\subsection{Single black hole}

To illustrate all of this, let us first examine the single centre solution. The harmonic function in the metric of (\ref{metric}) is now simply
\begin{equation}
H(r) = 1 + \frac{\mu}{r^{d-3}}.
\end{equation}
In this case, the solution is spherically symmetric. This additional symmetry implies that $\theta \equiv \Theta$ along $\gamma$. The constraint (\ref{constr}) reduces to
\begin{equation}
\label{rlambda}
 \frac{dr}{d\lambda} = H^{\frac{d-4}{d-3}}
\end{equation}
which can be integrated to give
\be
 r = \left[ \left( \mu^{1/(d-3)} + \lambda \right)^{d-3} - \mu \right]^{1/(d-3)}.
\ee
Expanding out for small $r$ we see that
\begin{equation}
\label{leadingr}
r \sim \left( (d-3) \mu^{(d-4)/(d-3)} \lambda \right)^{1/(d-3)}.
\end{equation}
Writing $H$ as a function of $\lambda$ gives
\be
 H = \frac{ \left( \mu^{1/(d-3)} + \lambda \right)^{d-3}}{ \left( \mu^{1/(d-3)} + \lambda \right)^{d-3} - \mu}.
\ee
Equation (\ref{tsol}) is now
\begin{equation}
t = v- \int H(\lambda)^2 d\lambda
\end{equation}
and hence
\be
\label{vdefsingle}
 dt = dv - H(\lambda)^2 d\lambda.
\ee
Finally we obtain the metric in the Gaussian null coordinate system:
\begin{equation}
\label{gaussnullmetric}
ds^2 = -H^{-2}dv^2+2dvd\lambda + \left(\mu^{1/(d-3)} + \lambda \right)^2 \left( d\Theta^2 + \sin^2 \Theta d\Omega_{d-3}^2 \right).
\end{equation}
The metric components are analytic functions of $\lambda$ so this defines an analytic extension of the solution through the horizon at $\lambda=0$ to negative values of $\lambda$. In this case, spherical symmetry implies that $g_{v\Theta}=0$ (and the Gaussian null coordinates coincide with Eddington-Finkelstein coordinates). However, the multi-centre solution will take the more general form (\ref{generalmetric}). 

\subsection{Strategy for multi-black holes}

To find the functions $r(\lambda,\Theta)$ and $\theta(\lambda,\Theta)$ for the multi-centre solution we substitute the power series expansions
\be
\begin{split}
\label{expansions}
r &= \sum_{m=1} c_m(\Theta,h_n) \lambda^{m/(d-3)} \\
\theta &= \Theta + \sum_{m=1} b_m(\Theta,h_n) \lambda^{m/(d-3)}
\end{split}
\ee
into equations (\ref{constr}) and (\ref{req}). The motivation for the choice of fractional powers in these expansions comes from the leading order behaviour of $r$ as a function of $\lambda$ for a single black hole, given in equation (\ref{leadingr}). The coefficients in these expansions are determined using computer algebra.

\section{Multi-centre solution in five dimensions}
\label{5dmulticentre}

\subsection{Determining the geodesic}

We begin by considering the $d=5$ solution, for which we can write the harmonics in the form
\begin{equation}
Y_n(\cos \theta) = \frac{\sin \left( (n+1) \theta \right)}{\sin \theta}.
\end{equation}
Upon solving the geodesic equation with the expansions (\ref{expansions}) as described in the previous section, and imposing the intial conditions (\ref{init}), we find the functions $r(\lambda,\Theta)$ and $\theta(\lambda,\Theta)$. The first few terms are
\be
\begin{split}
r &= \sqrt{2} \mu_1^{1/4} \lambda^{1/2} + \frac{h_0}{2\sqrt{2}\mu_1^{1/4}} \lambda^{3/2} + \mathcal{O}(\lambda^2), \\
\theta &= \Theta - \frac{2\sqrt{2} h_1 \sin \Theta}{\mu_1^{1/4}}\lambda^{3/2} + \mathcal{O}(\lambda^2).
\end{split}
\ee
More details, and further terms in the expansion, are given in Appendix A. The function $T(\lambda,\Theta)$ of equation (\ref{tsol}) is also given there.

\subsection{The metric is not $C^3$}

To investigate the smoothness of the metric, we consider the area of the $2$-sphere orbits of the $SO(3)$ symmetry in the geometry, which is a scalar invariant of the solution. The metric on these $S^2$'s is obtained by restricting the full metric to the space spanned by the $SO(3)$ Killing fields. Since the Killing fields have the same differentiability as the full metric, it follows that the metric on these $S^2$'s must also have the same differentiability as the full metric. The area of $S^2$ is
\begin{equation}
A_2 \equiv Hr^2 \sin^2 \theta.
\end{equation}
We can determine how $A_2$ varies along the null geodesic $\gamma$ as it approaches the horizon by 
substituting our expansions for $r$ and $\theta$ along the geodesic into this expression. This gives
\begin{multline}
A_2 = \mu_1 \sin^2 \Theta + 2 h_0 \sqrt{\mu_1} \sin^2 \Theta \lambda + \left( h_0^2 - 4 h_2 \mu_1 \sin^2 \Theta \right) \sin^2 \Theta \lambda^2 \\ - \frac{128 \sqrt{2}}{5} h_3 \mu_1^{5/4} \sin^2 \Theta \cos \Theta \lambda^{5/2} + \mathcal{O}(\lambda^3).
\end{multline}
The presence of a term of $\mathcal{O}(\lambda^{5/2})$ indicates that this quantity is not $C^3$ at $\lambda=0$, and we infer that the metric cannot be $C^3$ at the horizon unless $h_3$ vanishes. However, the explicit expression for $h_n$ (\ref{hnalld}) shows that $h_3 \ne 0$ for a 2-centre solution. If there are more than 2 centres then $h_3$ is still non-vanishing unless the parameters are finely tuned to make $h_3$ vanish. We conclude that generic multi-centre solutions do not admit $C^3$ horizons.

\subsection{Transforming the metric}

We now want to construct a coordinate system in a neighbourhood of the horizon in which the metric is $C^2$. The first step is to convert the metric (\ref{metric}) outside the horizon to Gaussian null coordinates as explained above. The fact that the metric is not $C^3$ at the horizon will be reflected in the presence of ${\cal O}(\lambda^{5/2})$ terms in the metric. 

In practice, $r(\lambda,\Theta)$ and $\theta(\lambda,\Theta)$ are expressed as infinite expansions in $\lambda$. In order to be as explicit as possible about the coordinate transformation, we shall truncate these expansions, keeping just as many terms as are required to make the metric coefficients $C^2$ functions of $\lambda$. This amounts to keeping just the terms written out explicitly in Appendix A. Of course, by performing this truncation we are no longer dealing with exact Gaussian null coordinates as defined above, but with "nearly Gaussian null coordinates", in which the metric differs from the Gaussian null form (\ref{generalmetric}) by terms of order $\lambda^{5/2}$. For example, there will be non-vanishing components $g_{\lambda \lambda}$ and $g_{\lambda \Theta}$ of order $\lambda^{5/2}$.  

The multi-centre $d=5$ metric in these coordinates in a neighbourhood of the black hole horizon at $r=0$ is
\begin{multline}
\label{exterior}
ds^2 = \left( -\frac{4}{\mu_1}\lambda^2 + \frac{12h_0}{\mu_1^{3/2}}\lambda^3 + \mathcal{O}(\lambda^{7/2}) \right) dv^2 + (2 + \mathcal{O}(\lambda^{9/2}))d\lambda dv \\ + \mathcal{O}(\lambda^{5/2})d\lambda^2 + \mathcal{O}(\lambda^{5/2})d\lambda d\Theta + \mathcal{O}(\lambda^{5/2})dvd\Theta \\ + \left( \mu_1 + 2\sqrt{\mu_1}h_0 \lambda + \left( h_0^2 + 8 h_2 \mu_1 \sin^2 \Theta \right) \lambda^2 +\mathcal{O}(\lambda^{5/2}) \right) d\Theta^2 \\ + \left( \mu_1 \sin^2 \Theta + 2\sqrt{\mu_1}h_0\sin^2 \Theta \lambda + \left( h_0^2 - 4h_2 \mu_1 \sin^2 \Theta \right) \sin^2 \Theta \lambda^2 + \mathcal{O}(\lambda^{5/2}) \right) d\Omega_2^2
\end{multline}
The metric is not well-defined at $\Theta = 0, \pi$. However, there is no indication in the above expansion that these are anything other than the usual coordinate singularities of spherical polar coordinates.

\subsection{Extending through the horizon}

\label{extend}

We will extend the solution through the horizon as follows. We assume that the interior metric takes the form
(\ref{metric}) with the same coordinate ranges (e.g. $r>0$) but with a {\it different} harmonic function $\hat{H}(r,\theta)$:
\begin{equation}
\label{interiormetric}
ds^2 = -\hat{H}^{-2} dt^2 + \hat{H}(dr^2 + r^2 d\theta^2 + r^2 \sin^2 \theta d\Omega_2^2),
\end{equation}
where $\hat{H}$ is chosen to agree with $H$ at leading order:
\begin{equation}
\hat{H} = \frac{\mu_1}{r^2} + \sum_{n=0}^{\infty} \hat{h}_n r^n Y_n (\cos \theta).
\end{equation}
Recall that we are assuming that the Killing field $V$ that generates time translations can be extended through the horizon. We choose the time-orientation of the interior solution so that $V=\partial/\partial t$.

We assume that the interior Maxwell field is as in (\ref{maxwell}) but with a sign change (which obviously still gives a solution):
\be
\label{interiormaxwell}
 A = \hat{H}^{-1} dt.
\ee
We will convert the interior metric to the "nearly Gaussian null" form, and then attempt to match it onto the nearly Gaussian null form of the exterior metric given above. This will impose some restrictions on the coefficients $\hat{h}_n$.

In the exterior region, $\lambda>0$ and $\lambda$ is the affine parameter along a {\it past}-directed geodesic $\gamma$. It is convenient to define a parameter $\hat{\lambda}$ in the interior region to be the affine parameter along a geodesic $\hat{\gamma}$ that obeys all the same conditions as $\gamma$ except that $\hat{\gamma}$ is {\it future}-directed (which amounts to replacing the condition $U \cdot V=1$ with $U \cdot V = -1$ above). We then construct Gaussian null coordinates as before. The only difference is that we will have $g_{v\hat{\lambda}}=-1$ whereas $g_{v\lambda}=+1$. To determine $\hat{\gamma}$ we proceed as before. The equations are exactly the same except for a sign change in (\ref{energycons}) because the geodesic is now future directed. This does not affect the equations governing $r$ and $\theta$ hence the expansions for $r$ and $\theta$ are exactly the same as before but with $\lambda \rightarrow \hat{\lambda}$ and $h_n \rightarrow \hat{h}_n$. Hence we can define nearly Gaussian null coordinates as before, with the coordinate $v$ defined by
\be
 t = v +  \hat{T}(\hat{\lambda},\Theta), \qquad \hat{T}(\hat{\lambda},\Theta) \equiv \int \hat{H}(\hat{\lambda},\Theta)^2 d\hat{\lambda}.
\ee
We can determine $\hat{T}(\hat{\lambda},\Theta)$ by replacing $\lambda \rightarrow \hat{\lambda}$ and $h_n \rightarrow \hat{h}_n$ in $T(\lambda,\Theta)$. The metric in these coordinates is obtained from the exterior metric (\ref{exterior}) by the same replacements and the sign change $g_{v\lambda}=1 \rightarrow g_{v\hat{\lambda}}=-1$:
\begin{multline}
\label{interior}
ds^2 = \left( -\frac{4}{\mu_1}\hat{\lambda}^2 + \frac{12\hat{h}_0}{\mu_1^{3/2}}\hat{\lambda}^3 + \mathcal{O}(\hat{\lambda}^{7/2}) \right) dv^2 + (-2 + \mathcal{O}(\hat{\lambda}^{9/2}))d\hat{\lambda} dv \\ + \mathcal{O}(\hat{\lambda}^{5/2})d\hat{\lambda}^2 + \mathcal{O}(\hat{\lambda}^{5/2})d\hat{\lambda} d\Theta + \mathcal{O}(\hat{\lambda}^{5/2})dvd\Theta \\ + \left( \mu_1 + 2\sqrt{\mu_1}\hat{h}_0 \hat{\lambda} + \left( \hat{h}_0^2 + 8 \hat{h}_2 \mu_1 \sin^2 \Theta \right) \hat{\lambda}^2 +\mathcal{O}(\hat{\lambda}^{5/2}) \right) d\Theta^2 \\ + \left( \mu_1 \sin^2 \Theta + 2\sqrt{\mu_1}\hat{h}_0\sin^2 \Theta \hat{\lambda} + \left( \hat{h}_0^2 - 4\hat{h}_2 \mu_1 \sin^2 \Theta \right) \sin^2 \Theta \hat{\lambda}^2 + \mathcal{O}(\hat{\lambda}^{5/2}) \right) d\Omega_2^2
\end{multline} 
Finally, we define the coordinate $\lambda$ for the interior region by $\lambda=-\hat{\lambda}$. It is then clear that the interior metric (\ref{interior}) for $\lambda<0$ matches onto to the exterior metric (\ref{exterior}) for $\lambda>0$ in a $C^2$ manner (i.e. up to ${\cal O}(|\lambda|^{5/2})$) provided that
\begin{equation}
\label{relations}
\hat{h}_0 = - h_0 \quad \quad \hat{h}_2 = h_2.
\end{equation}
The other multipole moments $\hat{h}_n$ are unconstrained, hence infinitely many interior solutions can be matched onto a given exterior solution so that the metric is $C^2$ at the horizon. Somewhat surprisingly, $\hat{h}_1$ is unconstrained.

\subsection{The Maxwell field}

We may now check the degree of differentiability of the Maxwell field. In the exterior region, the potential in the nearly Gaussian-null coordinates is
\begin{equation}
A = -H^{-1} dt = -H^{-1} (dv - \partial_{\lambda}T d\lambda - \partial_{\Theta}T d\Theta).
\end{equation}
Expanded out for small $\lambda$, this is
\begin{multline}
A = \left( -\frac{2}{\sqrt{\mu_1}} \lambda + \frac{3h_0}{\mu_1}\lambda^2 + \frac{32\sqrt{2}h_1 \cos \Theta}{5 \mu_1^{3/4}}\lambda^{5/2} + \mathcal{O}(\lambda^3) \right) dv \\ + \left( \frac{\sqrt{2}}{\mu_1} \lambda^{-1} + \frac{3h_0}{4} + \frac{8\sqrt{2}}{5} \mu_1^{1/4} h_1 \cos \Theta \lambda^{1/2} +{\cal O}(\lambda) \right) d\lambda \\ - \left(\frac{32\sqrt{2}}{5}\mu_1^{1/4} h_1 \sin \Theta \lambda^{3/2} + \mathcal{O}(\lambda^2) \right) d\Theta.
\end{multline}
The singular ${\cal O}(\lambda^{-1})$ term and the constant term in $A_\lambda$ are obviously pure gauge. 
In the interior region we have
\be
 A = \hat{H}^{-1} dt = H^{-1} (dv + \partial_{\hat{\lambda}}\hat{T} d\hat{\lambda} + \partial_{\Theta}\hat{T} d\Theta)
\ee
which gives
\begin{multline}
A = \left( \frac{2}{\sqrt{\mu_1}} \hat{\lambda} - \frac{3\hat{h}_0}{\mu_1}\hat{\lambda}^2 - \frac{32\sqrt{2}\hat{h}_1 \cos \Theta}{5 \mu_1^{3/4}}\hat{\lambda}^{5/2} + \mathcal{O}(\hat{\lambda}^3) \right) dv \\ + \left( \frac{\sqrt{2}}{\mu_1} \hat{\lambda}^{-1} + \frac{3\hat{h}_0}{4} + \frac{8\sqrt{2}}{5} \mu_1^{1/4} \hat{h}_1 \cos \Theta \hat{\lambda}^{1/2} +{\cal O}(\hat{\lambda}) \right) d\hat{\lambda} \\ - \left(\frac{32\sqrt{2}}{5}\mu_1^{1/4} \hat{h}_1 \sin \Theta \hat{\lambda}^{3/2} + \mathcal{O}(\hat{\lambda}^2) \right) d\Theta
\end{multline}
Upon defining $\lambda = -\hat{\lambda}$ for the interior region it is clear that the Maxwell field strength is $C^0$ in these coordinates. However, it is not $C^1$ because $F_{\lambda \Theta} = {\cal O}(|\lambda|^{1/2})$ unless the solution is finely tuned so that $h_1 = 0$ and we choose $\hat{h}_1=0$ (and $\hat{h}_0=-h_0$ as in (\ref{relations})), in which case the Maxwell field strength becomes $C^1$.

\subsection{Smoothness enhancement}
\label{balanced}

We have seen that the metric of a $d=5$ multi-black hole solution is generically $C^2$ but not $C^3$ at the horizon, and the Maxwell field strength is generically $C^0$ but not $C^1$. However, we have commented that the differentiability of the metric can be increased if $h_3=0$ and that of the Maxwell field increases if $h_1=0$. This increase in differentiability for certain configurations was also discussed in \cite{Welch95} for the particular case of a three-black hole solution, where the masses and locations of the black holes can be chosen in such a way that the central black hole horizon may be made more differentiable. A similar effect had been observed previously for cosmological multi-black hole solutions in $d=4$ \cite{Brill94}.

In general, the lack of smoothness arises from the fact that $r \sim \lambda^{1/2}$, so odd powers of $r$ are not smooth functions of $\lambda$. Hence the degree of differentiability increases if we arrange for the lowest odd multipole moments in the expansion of $H$ to vanish. If we we arrange the black holes so that {\it all} odd multipole moments vanish then the horizon (of the black hole at $r=0$) becomes smooth, in fact analytic. We shall now discuss this case in more detail.

Assume that the sources are arranged in such a way that there is a reflection symmetry in the plane $\theta=\pi/2$. In other words we have $N=2M+1$ black holes with parameters $(\mu_1,0)$, $(\mu_i,\pm a_i)$, $i=2\ldots M+1$. Then $h_{2n+1}=0$ for all $n$. We write $H$ has
\be
H = H_0 + \bar{H}, \qquad H_0 = \frac{\mu_1}{r^2} + h_0, \qquad \bar{H} = \sum_{n=1}^{\infty} h_{2n} r^{2n} Y_{2n}(\cos \theta).
\ee
We can now construct an analytic extension through $r=0$ using essentially the same method as in $d=4$ \cite{Hawking}. Define $\lambda>0$ by (compare (\ref{rlambda}))
\be
 \frac{dr}{d\lambda} = H_0^{1/2},
\ee
which gives
\be
 r = \left(2 \mu_1^{1/2} \lambda + h_0 \lambda^2\right)^{1/2}.
\ee
Define $v$ by (compare (\ref{vdefsingle}))
\be
\label{vdefbalanced}
 dt = dv - H_0(\lambda)^2 d\lambda.
\ee
The solution in coordinates $(v,\lambda,\theta,\Omega_2)$ is then analytic in $\lambda$ at $\lambda=0$ and can therefore be extended through the horizon to negative values of $\lambda$. In the interior of the black hole we can define a new coordinate $r$ as 
\be
 r = \left(-2 \mu_1^{1/2} \lambda - h_0 \lambda^2\right)^{1/2}.
\ee
and a new coordinate $t$ by (\ref{vdefbalanced}) to bring the interior solution to the form (\ref{interiormetric},\ref{interiormaxwell}) with a new harmonic 
function
\be
\hat{H}(r,\theta) \equiv - H(ir,\theta).
\ee  
Note that this obviously obeys the conditions (\ref{relations}) required for the extension to be $C^2$, but now we see that analyticity uniquely determines all of the higher coefficients $\hat{h}_{2n}$ too.

This solution is analytic at the horizon of the black hole at $r=0$. However, it will not be smooth at the horizons of the other black holes. It is interesting to ask whether one can construct a multi-black hole solution for which {\it all} black holes have analytic horizons. The above analysis, reveals that this will be the case if the sources are reflection-symmetric about every black hole on the axis. It is easy to see that this implies that there must be infinitely many black holes present, equally spaced on the axis of symmetry, with masses alternating between two values $\mu_1$ and $\mu_2$. Such a solution with $\mu_1=\mu_2$ was studied in \cite{Myers}, where a periodic identification was imposed in order to obtain a solution describing a single black hole localized on a Kaluza-Klein circle. If $\mu_1 \ne \mu_2$ then one can periodically identify in order to obtain a solution describing two black holes localized at antipodal points of a Kaluza-Klein circle. (Obviously one can also construct solutions for which the two black holes are localized anywhere on the circle but the horizons will be analytic only when the black holes are at antipodal points.)

\section{Multi-centre solutions in $d > 5$}

\label{higherdsmoothness}

Having warmed up with the case $d=5$, we now proceed to look at the smoothness for $d > 5$. 
Using the power series ansatz given in section \ref{start}, we can solve the geodesic equation order-by-order as before, with $d$ allowed to take any integer value greater than $5$. Upon substitution into the geodesic equation (\ref{req}), with the null constraint (\ref{constr}) and the initial conditions (\ref{init}) imposed as before we find
\begin{equation}
\begin{split}
c_2 &= c_3 = \ldots = c_{d-3} = 0 \\
b_1 &= b_2 = \ldots = b_{d-3} = 0
\end{split}
\end{equation}
and the first few non-zero terms in the expansion are
\begin{equation}
\label{generalcoeffs}
\begin{split}
c_1 &= (d-3)^{\frac{1}{d-3}} \mu_1^{\frac{d-4}{(d-3)^2}} \\
c_{d-2} &= \frac{(d-4)}{2}(d-3)^{\frac{4-d}{d-3}} \mu_1^{-\frac{1}{(d-3)^2}} h_0\\
c_{d-1} &= \frac{(d-4)(d-3)^{\frac{d-1}{d-3}}}{2d-5} \mu_1^{\frac{d-5}{(d-3)^2}} h_1 \cos \Theta \\
b_{d-2} &= -(d-3)^{\frac{d-2}{d-3}} \mu_1^{-\frac{1}{(d-3)^2}} h_1 \sin \Theta \\
b_{d-1} &= \frac{d-2}{4}(d-3)^{\frac{d-1}{d-3}} \mu_1^{\frac{d-5}{(d-3)^2}} h_2 \sin (2\Theta).
\end{split}
\end{equation}
With these we can check the smoothness of the area of the $S^{d-3}$ along the geodesic. The area is
\begin{equation}
A_{d-3} \equiv H^{\frac{2}{d-3}} r^2 \sin^2 \theta
\end{equation}
which expands out for small $\lambda$ to give
\begin{multline}
A_{d-3} = \mu_1^{\frac{2}{d-3}} \sin^2 \Theta + 2 \mu_1^{\frac{1}{d-3}} h_0 \sin^2 \Theta \lambda + \mu_1^{\frac{3d-11}{(d-3)^2}} (d-3)^{\frac{d-1}{d-3}} h_2 \sin^4 \Theta \lambda^{\left(1+\frac{2}{d-3}\right)} \\ + \mathcal{O}(\lambda^{\left(1+\frac{3}{d-3}\right)}),
\end{multline}
where we have neglected terms of order $\mathcal{O}(\lambda^{\frac{2d-6}{d-3}})$ as these are only of the same order as $\lambda^{(1+2/(d-3))}$ for $d = 5$. 

For $d=5$ the term of order $\lambda^{1+2/(d-3)}$ is smooth so the first term containing a non-integer power of $\lambda$ is of order $\lambda^{\frac{5}{2}}$, which implies that the metric is not $C^3$ as we saw above. However, if $d > 5$ then the term of order $\lambda^{1+2/(d-3)}$ does spoil smoothness, and implies that
the metric of a $d>5$ multi-centre solution is not $C^2$ at the horizon. Note that it is {\it not} possible to arrange the black holes so that the coefficient $h_2$ vanishes (equation (\ref{hnalld})) hence there is no analogue of the "smoothness enhancement" that is possible in $d=5$.

Using the same procedure as that used for $d=5$, we can find the expansions (\ref{expansions}) for $d > 5$. With these it is a straightforward task to introduce Gaussian null coordinates as described above. The solution in these coordinates is given in Appendix B. The extension through the horizon is performed in exactly the same manner as for $d=5$ (section \ref{extend}). Again we assume that the interior solution is described by a new harmonic function $\hat{H}(r,\theta)$ and Maxwell potential given by
\be
\begin{split}
\hat{H} &= \frac{\mu_1}{r^{d-3}} + \sum_{n=0}^{\infty} \hat{h}_n r^n Y_n (\cos \theta) \\
A &= \hat{H}^{-1}dt
\end{split}
\ee
and then we determine the restrictions on $\hat{h}$ by matching this interior solution (expressed in Gaussian null coordinates) with the exterior one given in Appendix B. The analysis goes precisely as described in section \ref{extend}. We have already seen that a $C^2$ extension of the metric does not exist, so the best we can obtain is a $C^1$ extension, which imposes the condition
\be
\hat{h}_0 = -h_0.
\ee
The higher order coefficients $h_1$ and $h_2$ appear in the metric above order $\lambda$ and so are not fixed by the requirement of matching the solutions in a $C^1$ manner. The Maxwell field strength in these coordinates is $C^0$ but not $C^1$.

The fact that the metric is not $C^2$ suggests that there may be a curvature singularity at the horizon. To see that this is indeed the case, note that 
some components of the Riemann tensor in the Gaussian-null coordinates diverge at the horizon. For example.
\begin{equation}
R_{\lambda \Theta \Theta v} = \frac{2(\partial_{\lambda}^2 g_{\Theta \Theta}) g_{\Theta \Theta} - (\partial_{\lambda} g_{\Theta \Theta})^2}{4g_{\Theta \Theta}}
\end{equation}
which expands out as
\begin{multline}
\label{riemann}
R_{\lambda \Theta \Theta v} = (d-1)(d-3)^{2/(d-3)}\mu_1^{\frac{3d-11}{(d-3)^2}} h_2 \sin^2 \Theta \lambda^{\frac{5-d}{d-3}} + \mathcal{O}(\lambda^{\frac{6-d}{d-3}}).
\end{multline}
This diverges on the horizon for $d>5$. By considering the Riemann tensor in a parallely propagated frame, we can show that this is a genuine curvature singularity. We construct frame vectors parallely propagated along a null geodesic with tangent vector $U=\partial/\partial \lambda$ as follows. Set $e^0=U$. Then, on the horizon, define $e^1=V$ and $e^2 = g_{\Theta\Theta}^{-1/2} \partial/\partial \Theta$ so that $e^0 \cdot e^1 = 1$, $e^0 \cdot e^2 = e^1 \cdot e^2 = 0$, $(e^0)^2=(e^1)^2=0$, $(e^2)^2=1$. Extend $e^{1,2}$ off the horizon by demanding that they are parallely propagated along the geodesic: $U \cdot \nabla e^{1,2}=0$. This preserves the orthogonality relations. Then, in this basis, we have 
\be
 R^{0221} = R_{abcd} e^{0a} e^{2b} e^{2c} e^{1d} = g_{\Theta\Theta}^{-1} R_{\lambda \Theta \Theta v} + \ldots
\ee
where the ellipsis denotes subleading terms.\footnote{Here we have used the fact that the leading non-smooth terms in the metric are ${\cal O}(\lambda^{(d-1)/(d-3)})$ and hence the leading non-smooth terms in the Riemann tensor in Gaussian null coordinates are ${\cal O}(\lambda^{(5-d)/(d-3)})$. Therefore no component of the Riemann tensor is more divergent than $R_{\lambda \Theta \Theta v}$.} From this we see that $R^{0221}$ diverges at the horizon so there is a parallely-propagated curvature singularity there.

\section{Discussion}

We should mention two potential loopholes in our work. First, we have considered multi-black hole solutions for which the black holes lie on a common axis. The exterior solution has $R \times SO(d-3)$ isometry group, and we have considered only extensions that preserve this symmetry. It is conceivable that smoother extensions exist, but break some of the symmetry. However, for $d>5$, this cannot be the case because we have shown that the solution has a curvature singularity at the horizon, and hence the metric cannot be $C^2$ there for any extension. For $d=5$, we expect that a similar argument could be used to prove that some components of the curvature tensor in a parallely propagated frame must have divergent derivatives at the horizon, proving that no $C^3$ extension of the metric exists.  This would require constructing the parallely propagated frame beyond leading order. However, a simpler argument shows that the Maxwell tensor cannot be extended in a $C^1$ manner: the basis component $F^{02} = F_{\lambda a} e^{2a} = F_{\lambda \Theta} e^{2\Theta} = {\cal O}(\lambda^{1/2})$ is not $C^1$ along the geodesic (here we used $e^{2v}=0$ from $U \cdot e^2=0$).

Second, our extension through a horizon makes use of spherical polar coordinates, and therefore exhibits singularities at the poles, where the horizon intersects the axis of symmetry. There is no indication that these are anything other than coordinate singularities (the axis itself is regular between the black holes) but a more rigorous treatment would require investigation of this issue.

\bigskip

\centerline{\bf Acknowledgments}

\noindent GNC is supported by the University of Nottingham. HSR is a Royal Society University Research Fellow. We thank P. Chrusciel for useful discussions.

\appendix
\section{Expansions for $r$ and $\theta$ for $d=5$}

The full expansions for $r(\lambda,\Theta)$ and $\theta(\lambda,\Theta)$ are now given, up to the order required to eliminate from the metric all divergent terms and fractional powers up to $\lambda^{5/2}$. The coefficients in these expansions were obtained by substituting the ansatz (\ref{expansions}) into equations (\ref{constr}), (\ref{req}) and solving order by order. Note that the initial condition $d\theta/d\lambda=0$ implies that $b_2=0$. It is convenient to relax this condition, and take the limit $b_2 \rightarrow 0$ at the end of the calculation. This is because the equations exhibit degeneracy when $b_2=0$. For example, $b_3$ is determined by plugging the expansions into (\ref{constr}) and examining the coefficient of $\lambda^{1/2}$, which gives an equation of the form $b_2 (b_3 - \ldots)=0$. Hence if $b_2 \ne 0$ then this determines $b_3$. However, if $b_2 =0$ then one has to go to higher order to determine $b_3$ (with the same result).

\begin{multline}
r = \sqrt{2}\mu_1^{1/4} \lambda^{1/2} + \frac{h_0}{2\sqrt{2}\mu_1^{1/4}}\lambda^{3/2} + \frac{4}{5} h_1 \cos \Theta \lambda^2 + \frac{32 h_2 \mu_1 \csc \Theta \sin (3\Theta) - 3 h_0^2}{48 \sqrt{2} \mu_1^{3/4}}\lambda^{5/2} \\ + \frac{4(20 h_3 \mu_1 \cos (2\Theta) - h_0 h_1)\cos \Theta}{35\sqrt{\mu_1}}\lambda^3 \\ + \frac{25h_0^3 + 64\mu_1((86 \cos (2\Theta) -89)h_1^2 + 25(1+2\cos(2\Theta) + 2 \cos (4\Theta))h_4\mu_1)}{1600\sqrt{2}\mu_1^{5/4}}\lambda^{7/2} \\ + \frac{(3h_0^2h_1 + 24h_0 h_3 \mu_1 \cos(2\Theta) + 28 \mu_1 ((4+8\cos(4\Theta))h_5\mu_1 - 50 h_1 h_2 \sin^2 \Theta))\cos\Theta}{63\mu_1}\lambda^4 \\ + \frac{1}{1612800\sqrt{2}\mu_1^{7/4}}\left( 33600(1+2\cos(2\Theta))h_0^2 h_2 \mu_1 - 7875h_0^4 \right. \\ + 2304h_0 \mu_1 ((6683-6610\cos(2\Theta))h_1^2 + 245(1+2\cos(2\Theta) + 2\cos(4\Theta))h_4\mu_1) \\ + 1024\mu_1^2((140\cos(2\Theta)-7455+7630\cos(4\Theta))h_2^2 + 72((194\cos(4\Theta) -116\cos(2\Theta) -58)h_1h_3 \\ \left. + 35(1+2\cos(2\Theta) + 2\cos(4\Theta) + 2\cos(6\Theta))h_6\mu_1)) \right) \lambda^{9/2}
\end{multline}
\begin{multline}
\theta = \Theta - \frac{2\sqrt{2} h_1 \sin \Theta}{\mu_1^{1/4}}\lambda^{3/2} - 3 h_2 \sin (2\Theta) \lambda^2 \\ + \frac{\sqrt{2}}{10\mu_1^{3/4}} \left( 17 h_0 h_1 \sin \Theta - 8 h_3 \mu_1 \sin \Theta - 24 h_3 \mu_1 \sin (3\Theta) \right) \lambda^{5/2} \\ - \frac{2}{5\sqrt{\mu_1}} \left( 5 h_4 \mu_1 \sin (2\Theta) + 10 h_4 \mu_1 \sin (4\Theta) - 9 h_1^2 \sin (2\Theta) - 5 h_0 h_2 \sin (2\Theta) \right) \lambda^3
\end{multline}
\begin{multline}
T = -\frac{\mu_1}{4}\lambda^{-1} + \frac{3}{4}h_0\sqrt{\mu_1} \ln \lambda + k + \frac{16\sqrt{2}}{5} h_1 \mu_1^{3/4} \cos \Theta \lambda^{1/2} \\ + \left( \frac{11h_0^2}{16} + \frac{5}{3} h_2 \mu_1 \csc \Theta \sin (3\Theta) \right) \lambda + \frac{4\sqrt{2}}{35} \mu_1^{1/4} (19h_0 h_1 + 40 h_3 \mu_1 \cos (2\Theta)) \cos \Theta \lambda^{3/2} \\ + \frac{1}{400\sqrt{\mu_1}} \left( 25h_0^3 + 750(1+2\cos (2\Theta))h_0 h_2 \mu_1 + 4 \mu_1 ((469-106 \cos (2\Theta))h_1^2 \right. \\ \left. +175(1+2\cos(2\Theta)+2\cos(4\Theta))h_4\mu_1 ) \right) \lambda^2 + \frac{1}{315\sqrt{2}\mu_1^{1/4}} \left( (345h_0^2h_1 \right. \\ \left. + 4272 h_0 h_3 \mu_1 \cos (2\Theta) + 224 \mu_1 ((40-13\cos(2\Theta))h_1h_2 + 8(1+2\cos (4\Theta))h_5 \mu_1))\cos \Theta \right) \lambda^{5/2} \\+ \frac{1}{100800\mu_1} \left( 71400(1+2\cos (2\Theta))h_0^2h_2\mu_1 - 1575h_0^4 + 144 h_0\mu_1((943+866\cos(2\Theta))h_1^2 \right. \\ \left. + 2205(1+2\cos(2\Theta)+2\cos(4\Theta))h_4\mu_1) + 256\mu_1^2(70(51+32\cos(2\Theta) - 11\cos(4\Theta))h_2^2 \right. \\ \left. + 9((416+832\cos(2\Theta) - 228\cos(4\Theta))h_1h_3 + 105(1+2\cos(2\Theta)+2\cos(4\Theta)+2\cos(6\Theta))h_6\mu_1))\right) \lambda^3
\end{multline}
with $k$ an arbitrary constant. 

\section{Gaussian null coordinates for $d>5$}

The metric for $d > 5$ in the Gaussian null coordinate system is given by
\begin{multline}
ds^2 = \left(-\frac{(d-3)^2}{\mu_1^{2/(d-3)}} \lambda^2 + \frac{(d-2)(d-3)^2h_0}{\mu_1^{3/(d-3)}} \lambda^3 + \frac{2(d-1)(d-3)^{\frac{4d-11}{d-3}}}{(2d-5)\mu_1^{(2d-5)/(d-3)^2}} h_1 \cos \Theta \lambda^{\frac{3d-8}{d-3}} + \mathcal{O}(\lambda^{\frac{3d-7}{d-3}}) \right) dv^2 \\ + 2dvd\lambda + \left( \frac{8(1-d)(d-3)^{\frac{2d-5}{d-3}}}{(2d-5)\mu_1^{1/(d-3)^2}} h_1 \sin \Theta \lambda^{\frac{2d-5}{d-3}} + \mathcal{O}(\lambda^{\frac{2d-4}{d-3}}) \right) dvd\Theta \\ + \left( \mu_1^{\frac{2}{d-3}} + 2\mu_1^{\frac{1}{d-3}} h_0 \lambda - (d-3)^{\frac{2d-4}{d-3}} \mu_1^{\frac{3d-11}{(d-3)^2}} h_2 \sin^2 \Theta \lambda^{\frac{d-1}{d-3}} + \mathcal{O}(\lambda^{\frac{d}{d-3}}) \right) d\Theta^2 \\ + \left( \mu_1^{\frac{2}{d-3}} \sin^2 \Theta + 2 \mu_1^{\frac{1}{d-3}} h_0 \sin^2 \Theta \lambda - \mu_1^{\frac{3d-11}{(d-3)^2}} (d-3)^{\frac{d-1}{d-3}} h_2 \sin^4 \Theta \lambda^{\frac{d-1}{d-3}} + \mathcal{O}(\lambda^{\frac{d}{d-3}}) \right) d\Omega_{d-3}.
\end{multline}
For simplicity we have assumed that the expansions for $r$ and $\theta$ have been taken to infinity, allowing complete elimination of the $g_{\lambda \lambda}$ and $g_{\lambda \Theta}$ components. The Maxwell potential is
\begin{multline}
A = - \left( (d-3)\mu_1^{-1/(d-3)} \lambda + \frac{(2-d)(d-3)}{2}\mu_1^{-2/(d-3)}h_0\lambda^2 \right. \\ \left. + \left( \frac{1-d}{2d-5} \right) (d-3)^{\frac{3d-8}{d-3}} \mu_1^{\frac{2-d}{(d-3)^2}} h_1 \cos \Theta \lambda^{\frac{2d-5}{d-3}} + \mathcal{O}(\lambda^{\frac{2d-4}{d-3}}) \right) dv \\ + \left( (d-3)^{-1} \mu_1^{1/(d-3)} \lambda^{-1} + \frac{d-2}{2(d-3)}h_0 \right. \\ \left. + \frac{d-1}{2d-5} (d-3)^{\frac{d-2}{d-3}} \mu_1^{\frac{d-4}{(d-3)^2}} h_1 \cos \Theta \lambda^{\frac{1}{d-3}} + \mathcal{O}(\lambda^{\frac{2}{d-3}}) \right) d\lambda \\ - \left(2\left( \frac{d-1}{2d-5} \right) (d-3)^{\frac{2d-5}{d-3}} \mu_1^{\frac{d-4}{(d-3)^2}} h_1 \sin \Theta \lambda^{\frac{d-2}{d-3}}  + \mathcal{O}(\lambda^{\frac{d-1}{d-3}}) \right) d\Theta.
\end{multline}
As in $d=5$, the singular and constant terms in $A_{\lambda}$ may be removed with a gauge transformation.

\end{document}